\documentclass[aps,preprint,showpacs,prl]{revtex4}
\usepackage{graphics}

\begin{document}

\title{\boldmath{Are Bulk Axions in Models with Extra, Large, Compact
Dimensions
Observable? }}

\author{Shmuel Nussinov}
\email{nussinov@post.tau.ac.il}
\affiliation{School of Physics and Astronomy, Tel Aviv University\\
Ramat-Aviv, Tel Aviv 69978, Israel}

\date{June 21, 2004}

\begin{abstract}
The high degeneracy of KK modes in models which have bulk axions
moving
in some extra, compact dimensions which are larger than
O (Angstrom)
strongly tightens the supernova upper bounds on the axion photon
coupling making axions practically unobservable.  Conversely,
discovering
axions directly or indirectly will exclude such models. These drastic
conclusions are avoided if the supernova and bounds from solar axion
searches are relaxed.
\end{abstract}

\maketitle

While axions have not been discovered to date they hold the
fascination of a broad scientific community in approximately 30 years
since the original suggestion.

Axions were motivated by particle physics. A new symmetry and
attendant fields were introduced in order to dynamically relax to zero
the $\theta$
parameter in the otherwise dangerous parity violation: $\theta
E(c)B(c)$ term in
the QCD langrangian. The axion emerged as the light ``pseudo" Goldstone
boson
in the spontaneous breaking of this (Peccei-Quinn) symmetry.  Despite
many
efforts no alternative explanation of the experimentally tiny $\theta$
was
found.  Axions are of much interest in astrophysics as coherent, cold,
dark
matter and possibly manifest in hot, dense environments and/or strong
magnetic fields.

A key role is played by the axion two-photon coupling:
\begin{equation}
 g(a,\gamma,\gamma)\cdot E({\rm em})\cdot B({\rm em})
\label{g} %{ Eq. 1}
\end{equation}
which generates spontaneous $a \rightarrow 2 \gamma$ decays, coherent
$a \rightarrow \gamma$
transitions in strong $E$ or  $B$ fields, and contributes, via virtual
$\gamma \rightarrow$ axion $\rightarrow \gamma$
transition, to vacuum bifringing of polarized light in strong $B $
fields.  Astrophysical considerations, alongside experimental searches,
limit the allowed axion mass $m$ and coupling.  In particular, the
(conservative!) limit:
\begin{equation}
 M = g(a,\gamma,\gamma)^{-1} >  10^{10} {\rm GeV}
\label{conservlimit} % { Eq.2}
\end{equation}
is required to avoid catastrophic cooling via (volume) emission of
axions---from red giants and in supernovae
collapse.\cite{Raffelt:1999tx}

In supersymmetric/string theories the scalar ``dilaton" field fixing
the gauge coupling $g$ is a natural chiral partner of the pseudoscalar
axion
associated with varying $\Theta$.  It is therefore natural---in
theories with $n$,
large, compact, extra dimensions (LCED) to have (i) both axion and
dilaton on the
thin ``Brane", along with the standard model particle, or (ii) both in
the
``Bulk".  The possible connection of dilatons with bulk gravity favors
option (ii).

In the following we will assume that axions are bulk fields in some
LCED model.  Details of the model  beyond  the dimensionality, $n$, of
the extra
space where the axions live and its scale $R$ are irrelevant to our
discussion.

The possible states of bulk axions correspond to an $n$ dimensional
lattice of Kaluza-Klein  (KK) ``momenta":
\begin{equation}
  ( k_1,k_2,..k_n), \;\;\;  k_i = \ell_i/R
\label{KKmomenta}  %  { eq. 3}
\end{equation}
$\ell_i$ = 0 or integer. These states have masses:
\begin{equation}
  m(KK(\ell_1,\ell_2...\ell_n) = (1/R)\cdot \sqrt{\ell_1^2 + \ell_2^2 +...\ell_n^2}
\label{KKmasses}    %    { Eq. 4}
\end{equation}
The number:
\begin{equation}
  N(KK)(T(a)) \sim (R\cdot T(a))^n
\label{KKnumber} %     { Eq. 5}
\end{equation}
of KK axionic states with mass $m(KK) < T(a)$, which can be excited in
hot environments of temperature $T(a)$, can be very large, and enhanced
emission of axions from the sun has been
considered.\cite{DPRZ}
For a {\it given} $g(a,\gamma,\gamma)= M^{-1}$, axion emission is
enhanced in
theories with LCED by the above large number of KK species:
$EF(a) \sim (RT(a))^n$.
For $n=2$ and $R =eV^{-1}$, the values in Ref. \cite{DPRZ}, the enhancement factor
EF
in the sun with $T({\rm core} (o)) \sim = T({\rm sun}) \sim$1.4 KeV is:
\begin{equation}
   EF({\rm sun}) \sim (T({\rm sun})\cdot R)^n \sim 2\cdot 10^6.
\label{EFsun} %   { Eq. 6}
\end{equation}

Unfortunately LCED scenarios {\it with bulk axions} drastically {\it
suppress}
solar axion emission and even more so the number of detected axions.
If solar (or any!) axions are ever detected, most LCED models are
ruled out.

The argument is simple and is implicit in some early papers.\cite{CTY}
While axions from the nearby sun are expected to dominate axions
from other astrophysical sources like red giants and supernovae cores,
the
latter are $\sim$ 10 and 10$^4$ hotter than the stellar core,
respectively.
Thus the enhancement factor in supernovae is much larger than in the
sun.
\begin{equation}
  EF({\rm supernova}) = (T({\rm supernova})\cdot R)^n
                      = 10^{4n}\cdot EF({\rm sun})
\label{EFsupernova} % { Eq. 7 }
\end{equation}

Next recall that the lower bound on $M$ of Eq.  (\ref{conservlimit})
was derived by
demanding that the
axionic luminosity (scaling like $M^{-2}$) in the standard {\it single}
light axion scenario be less than the total luminosity of the
collapsing core.
To ensure this in the present scenario we need to compensate
$EF({\rm supernova})$
by increasing $M^2$ by $EF({\rm supernova}) = (T(\rm{supernova}) \cdot R)^n$.
Hence, in models with bulk axions in LECD the bound on the axion electromagnetic
coupling is much stronger than that of  Eq.  (\ref{conservlimit}) :
\begin{equation}
{\rm new \, (LCED) \, bound \, on} \; M = {\rm old \, bound \, on} M
\cdot (T({\rm supernova})\cdot R)^{n/2}
\label{newbound} %    { Eq. 8}
\end{equation}

Recalling Eq. (\ref{EFsun}) we find that relative to standard single
axion
scenarios the number of {\it detected} solar axions, which is
proportional to $M^{-4}$, is
reduced by
\begin{equation}
 (T(\rm{supernova})\cdot R)^{2n}/(T({\rm sun})\cdot R)^n.
\label{Tsupernova}                   %   { Eq . 9 }
\end{equation}
The parameters of Eq. (\ref{EFsun})  and T(supernova) $\sim$ 10 MeV
then imply
a  $\sim 10^{22}$ reduction!

Axion effects in pure terrestrial laboratory experiments---say, by
contributing to vacuum bifringing of laser light (PVLAS)---are also
strongly
suppressed in LCED models with bulk axions relative to the ordinary one
axion
scenario.

The ``elipticity" angle $\Psi$ which would be generated in the
PVLAS\cite{PVLAS} experiment with
one axion of mass $m$ and electromagnetic coupling, $M^{-1}$, is given
by\cite{Maiani}
\begin{equation}
   \Psi \sim (BL/M)^2 \cdot (E(\gamma)/L \cdot m^2)
\label{PVLS}    %      Eq (10)
\end{equation}
where $B$ is the strong magnetic field applied and $L$ is the distance
along
which  the initial linearly polarized laser light has accumulated the
elipticity.  Strictly the last equation applies only in the regime
$m$(axion) $ < E(\gamma)$. Other intermediate states, such as $e^{(+)}
e^{(-)}$
pairs in QED and heavier axions also contribute but along shorter
coherence lengths
and much less efficiently.

In the LCED scenario we could have, when $\lambda(\gamma) < R$, many,
$\sim(E(\gamma)\cdot R)^n$, axions with masses smaller than
$E(\gamma)$.
The elipticity contributed by each of these axion KK models adds up
coherently and, hence, the effect is enhanced by  this factor.

However, again, as in the previous cases, the supernova bound
reverses the conclusions.  Any putative $(E(\gamma)\cdot R)^n$
enhancement is
vitiated by  the much more drastic decrease (by $(T(SN)\cdot R)^n)$ of
the $M^{-2}$
factor in Eq. (\ref{PVLS})
required in order to satisfy the supernova cooling bound. This yields a
net
suppression by $(T(SN)/E(\gamma))^n \sim 10^{8n}$, with $T(SN) \sim$ 10
MeV and
$E(\gamma) \sim 0.1$ eV.

All the above suppressions  are avoided if $R < 1/T$ (supernova) $\sim$
20 Fermi. The lightest KK excitation is then heavier than 10-30 MeV,
making the LCED models irrelevant for astrophysical axions and vice
versa astrophysical considerations limit axion physics in the same old
way as in the ordinary four-dimensional models.

The supernova bounds on $M$ could conceivably be relaxed by having
quick axion $\rightarrow$ photon conversion in the strong magnetic
fields.  This
would be resonantly enhanced by axion plasmon degeneracy.
This conversion may be modified and potentially accelerated if as in
a recently modified LCED scenario where KK momentum conservation in the
bulk is relaxed allowing for decays of heavier KK axions into lighter
ones.\cite{MNPL}

We conclude with a few comments:

(i) The above phenomenological, simplistic discussion suggests
that
bulk axions, if allowed to move in the same extra dimensions as
originally conceived for gravity, are constrained (by the supernova
cooling
bound) to have an extremely small coupling to photons. At a slightly
deeper
theoretical level such tiny couplings and $M \sim M$(Planck) $\sim
10^{19}$ Gev
is expected: the same ``dilution" by bulk to brane volume ratio
suppresses
both axion and graviton couplings.  Indeed, supernovae were used also
to limit
directly models LCED and  many gravitational KK
excitations.\cite{Hannestad:2003yd}

Our approach of using just KK repetitions of axions may be
oversimplified and one may want to have a more consistent higher
dimensional framework in which the $P\cdot Q$ scenario and axions have
to be
considered a priori.\cite{DDDG}

(ii)  Consider axions of mass $m$. If the axions' mass, $m \sim
T(a)$, a fraction
\begin{equation}
  F(b) \sim (V({\rm escape}(a))/c)^3
\label{Vescape}       %        { Eq. 11}
\end{equation}
of the axions, namely, those  emitted with velocities smaller than
the escape velocity, will gravitationally bind to the star in question.
For the sun $(V({\rm escape}) \sim$ 560 km/sec) and $m({\rm axion})
\sim$ 5 KeV,
this fraction is $10^{-7}$. \cite{DZ1},\cite{DZ2}
If the lifetime for $a \rightarrow \gamma +\gamma \; t({\rm axion})
\sim m^{-3}$
can be as short  as 10$^{20}$ seconds for $m \sim$10 KeV, and if
further solar axion
luminosity was  a  $\sim$ few percent of the total solar luminosity,
then radiative decays of the bound solar axions could generate the
x-rays observed
from the quiescent sun  and
resolve other longstanding puzzles. One axion-like particle with the
desired
mass and coupling yields  an x-ray sharply peaked at one energy
$(E \sim m/2)$---contrary to observation---and a framework with many
axions is required.

(iii)  More massive $m \sim$10 MeV axion-like particles can be
emitted from
supernova cores.  {\it If} it radiatively decays during times shorter
than hubble
time, then the bound on diffuse $\sim$ 10 MeV ``relic" gammas from all
past supernovae
implies that only a small fraction $f \sim 10^{-5} -10^{-6}$ of the
supernova energy
can then be emitted via such axions. Since the photon flux scales as
$M^{-4}$
this improves the ordinary supernova bound on $M$  by $f^{-(1/4)} \sim$
20-30.

(iv)  The original observation by DiLella and Zioutas\cite{ DZ1}
that some massive
KK  axions are gravitationally captured, applies to neutron stars.
Using Eq. (\ref{Vescape})  and $V$(escape)(SN)/c $\sim$ 1/3, we find that
$\sim$ 5\%
of all axions with mass $m \sim T$(SN) $\sim$ 10-30 MeV will be
gravitationally
trapped. The bound of Eq. (\ref{newbound})  allows axions
to carry 1/2 of the $\sim$ 3-4 10$^{53}$ ergs collapse energy.
The total energy/mass of the captured axions is: 10$^{51}$
ergs/10$^{30}$
grams. The radiative decays of such axions yield $\sim$ 10 MeV gammas
with a
luminosity
\begin{equation}
  L(\gamma, {\rm bound \, axions}) \sim 10^5/t({\rm decay})({\rm ergs/sec})
\label{luminosity}    % { Eq. 12}
\end{equation}
About 1/4 of the gammas hit the surface and could make old neutron
stars shine with high surface temperatures.

Using $m = m({\rm axion})$ = 30 MeV we find\cite{DPRZ}
 $t({\rm decay}) \sim  64\pi \cdot M^2/m^3 \sim (M/30 {\rm MeV})^2
\cdot 10^{-20}$ sec.
which, with the minimal $M$ in Eq. (\ref{conservlimit})   (10$^{10}$
GeV) and corresponding
bound (\ref{newbound}), yields:  $t({\rm decay}) \sim  (R \cdot$ 30
MeV)$^n \cdot 10^3$ sec.
For the parameters used above $(R \sim$ eV$^{-1}$ and $n=2$) we find
$t({\rm decay}) \sim 10^{18}$ sec,
The resulting luminosity in Eq. (\ref {luminosity}), $\sim$ 10\%
$L$(solar) leads to
observable radiation from old neutron stars.
The above effect scales as $\sim M^{-4}$. Thus using $M \sim 10^{12}$
GeV---just 100
times the most conservative lower bound---completely evades it.

(v)  A more subtle and possibly dramatic effect discussed
context of halo particles\cite{GN},\cite{GDR2N}
is the migration of the accumulated  particles to the center of the
star.  The latter then form a ``mini black hole" which, in turn,
``gobbles up" the whole star into a bigger black hole!

The bosonic axions cannot sustain the strong gravity of the very
dense $(\rho \sim 10^{15}$ gr/cm$^3$) core by Fermi-pressure.
However, the predominant axionic interactions with matter are not
elastic but rather axionic Compton or Primakoff effects.  Rather than
lose
energy via multiple elastic collisions, the axions convert into
photons and get absorbed preempting the black hole scenario.
Rather
the absorbed axions contribute to the heating up of the old, cold
neutron just
like the neutron star halo axions' spontaneous radiative decay
discussed above.

\subsection{Acknowledgement}

 I am much indebted to Konstantin Zioutas for many helpful
discussions.  This particular note was motivated by his works with
Di Lella and my interest in recent years in solar axions
by his suggestions at the early stages of the Solax and
Cast experiments.

\end{document}